\documentclass[journal]{IEEEtran}

\ifCLASSINFOpdf
   \usepackage[pdftex]{graphicx}
\else
\fi

\usepackage{graphicx}
\usepackage[cmex10]{amsmath}
\usepackage{cite}

\begin{document}

\title{Coherent WDM Systems Using in-Line Semiconductor Optical Amplifiers}

\author{Amirhossein~Ghazisaeidi 
        
\thanks{A. Ghazisaeidi is with Nokia Bell Labs, 
Paris-Saclay, Route de Villejust, 91620, Nozay, France
e-mail: (amirhossein.ghazisaeidi@nokia-bell-labs.com).}}

\markboth{October ~2017}
{Shell \MakeLowercase{\textit{et al.}}: }

\maketitle

\begin{abstract}
A theory of nonlinear signal propagation in multi-span wavelength division multiplexed coherent transmission systems that employ the semiconductor optical amplifier as in-line amplifiers is presented for the first time. The rigorous derivation, based on time-domain first-order perturbation theory, is developed in detail. The end result is the expressions for the signal-to-noise ratio of the received sampled photocurrent, including contributions from noise, fiber-induced nonlinear distortions, and amplified-induced nonlinear distortions.   
\end{abstract}


\IEEEpeerreviewmaketitle

\section{Introduction}

\IEEEPARstart{R}{cent} years have witnessed the combination of some advanced techniques, such as high cardinality shaped constellations, single-channel digital nonlinear compensation, and adaptive-rate capacity approaching forward error correction codes, to boost the spectral efficiency and the throughput of coherent wavelength division multiplexed (WDM) multi-span single-mode fiber-optic transmission systems.\cite{Ghazisaeidi2016:ECOC2016}-\cite{JXCai2017:ECOCPDP2017}. The spectral efficiency of fiber-optic transmission systems is fundamentally limited by fiber nonlinear impairments\cite{Agrawalbook,Essiambre2010:JLT2010}. Single-channel nonlinear compensation cannot mitigate nonlinear interference stemming from the adjacent channels. Multi-channel nonlinear compensation is prohibitively complex; moreover, in meshed networks the adjacent channels are added and dropped along the optical path of the channel of interest (COI); therefore, the received samples of the adjacent channel might not be adequate for the multi-channel nonlinear equalizer to efficiently mitigate nonlinear impairments. Even if the complexity is not a concern, and all the adjacent channels are perfectly known and jointly processed at the receiver, the nonlinear interaction between signal and spontaneous emission noise (ASE), an effect which will be referred to as NSNI in this work, sets the fundamental upper limit on the achievable spectral efficiency using the standard digital backpropagation (DBP) nonlinear equalizer. In \cite{Ghazisaeidi2017:JLT} we have studied those fundamental limits, and demonstrated that the maximum improvement in spectral efficiency of conventional coherent WDM systems using full-field DBP is about 50-60\%. 

Besides nonlinearity, the throughput of conventional WDM systems is also limited by the total bandwidth of erbium-doped fiber amplifiers (EDFA). Recent experiments using C+L dual-band EDFAs have an optical transmission bandwidth about 9~THz\cite{Ghazisaeidi2016:ECOC2016}-\cite{JXCai2017:ECOCPDP2017}. In order to extend the transmission bandwidth, all-Raman in-line amplification might be used\cite{Qian2011:OFC,Sano2012:OFC}, but it has its own drawbacks, most importantly the distributed nature of all-Raman amplification, which is hardly compatible with routed networks topology, and its low power efficiency. Note that even if all-Raman amplification is a viable option for \emph{in-line} amplifications, transmitter-side boosters and receiver-side pre-amplifiers are still EDFAs in \cite{Qian2011:OFC,Sano2012:OFC}. 

Another alternative to EDFA and Raman optical amplification schemes is semiconductor optical amplifiers (SOA)\cite{Connelly2007:Book}. It has been shown that SOAs can deliver amplification bandwidths up to 120~nm \cite{Akiyama2004:OFC}. The SOA consists of an electronically pumped millimeter-long active waveguide, and is an integrable device providing lumped gain, thus, it is an attractive potential candidate for replacing in-line amplifiers, boosters and pre-amplifiers in a multi-span repeated transmission system. However, light amplification by the SOA is a nonlinear process; the gain fluctuates dynamically, at the nanosecond scale, in response to input signal intensity fluctuations; moreover, the typical noise figure (NF) of the SOA is around 7~dB or higher, whereas that of  modern EDFAs is around to 4-5~dB. Due to its nonlinear behavior and high NF, the SOA has not been used as an in-line amplifier in the past. In fact, most of the versatile applications of the SOA are based on exploiting its nonlinear behavior in order to realize various optical signal processing functions \cite{Connelly2007:Book}.   

In September 2017, Renaudier \emph{et al.} have demonstrated the first coherent WDM transmission system employing custom-designed in-line ultra-wideband SOAs, with more than 100~nm continuous optical amplification band instead of EDFAs, with improved nonlinear and noise performance\cite{Renaudier2017:ECOCPDP2017}. This new result may pave the way for novel SOA-based coherent WDM systems with considerable throughput increase with respect to the present-day EDFA-amplified systems. That demonstration is the main motivation of the present work, where, we present for the first time, a rigorous theory of nonlinear signal propagation in coherent WDM multi-span systems based on in-line SOAs. Note that although a vast literature exist on modeling the SOA device \emph{per se}, as well as for various optical signal processing applications (cf. below for more details), transmission systems with in-line SOA amplifiers have not been studied in the literature due to the above mentioned lack of motivation. To the author's knowledge, the only exception is the numerical investigations in \cite{Djalal2013:ECOC2013}.

Nonlinear signal propagation in EDFA-based coherent WDM systems has been extensively studied\cite{Ghazisaeidi2017:JLT}, \cite{Mecozzi2012:JLT2012}-\cite{Johannisson2013:JLT2013}\footnote{This list is not exhaustive by far. For a detailed discussion of various approaches to modeling nonlinear signal propagation in coherent systems cf. the introduction section of \cite{Ghazisaeidi2017:JLT} and the references therein.}. Nonlinear propagation in Raman-amplified systems is addressed in \cite{Curri2013:OptXpress}. In particular, in \cite{Ghazisaeidi2017:JLT} we have presented a detailed demonstration of time-domain perturbative approach to nonlinear wave propagation in EDFA-based coherent systems, first proposed in \cite{Mecozzi2012:JLT2012}. In \cite{Ghazisaeidi2017:JLT}, both signal-signal (SS) and noise-signal (NS) nonlinear distortions have been rigorously studied, and the signal-to-noise ratio (SNR) of the received sampled signals are computed with and without nonlinear compensation. The departure point of the present work is \cite{Ghazisaeidi2017:JLT}. For the sake of simplicity, we limit our analysis to single-polarization fields and ignore the NSNI. We replace EDFAs by SOAs, and follow the lines of \cite{Ghazisaeidi2017:JLT}. We introduce necessary modifications to the theory, due to the presence of the gain dynamics of the SOAs, and compute the variance of total nonlinear SS distortions of the sampled received signal.

The main new ingredient of the present work is the SOA model. An extensive literature on modeling the behavior of the SOA exists (cf. \cite{Agrawal1989:JQE}-\cite{Antonelli2016:JLT} and the references therein for a non-exhaustive literature review). On one extreme there exist very sophisticated space-resolved models with tens of parameters that allow for accurate modeling most of the optical and electrical physical processes inside the SOA, \emph{e.g}, the space-dependence of material gain and optical field inside the SOA waveguide, forward and backward propagation, frequency dependence of gain and other parameters, various radiative radiative and non-radiative recombination processes \emph{etc.}, and are suitable for in-depth studies of the device physics\cite{Connelly2007:JQE}. On the other extreme, there are very simple reservoir models, containing only few parameters, yet successfully capturing the main features of the gain dynamics of the SOA, and so are more useful for system studies where SOAs are combined with many other components in a system architecture \cite{Agrawal1989:JQE}. In past, we have used the reservoir model of\cite{Agrawal1989:JQE} to study the statistical properties of SOA nonlinear noise\cite{Ghazisaeidi2009:JLT2009}-\cite{Ghazisaeidi2010:PTL}, the nonlinear patterning effect in the SOA\cite{Ghazisaeidi2010:JQE}, and digital post-compensation of SOA nonlinearity \cite{Ghazisaeidi2011:JLT,Siamak2013:OFC}, and have found satisfactory match with experimental data. In this paper we adopt the model presented in \cite{Agrawal1989:JQE} for the in-line  SOAs in the multi-span coherent WDM systems, and derive expressions for the SNR of the sampled received signal, using the time-domain first-order regular perturbation theory (FRP), following the steps of\cite{Ghazisaeidi2017:JLT}.

The structure of the paper is as follows. In Section II we present the theory. Section II.A is devoted the the basic definitions and problem statement, In section II.B the SOA model is presented, In section II.C. The perturbative analysis of the gain dynamics is presented in II.D. In section II.E the full perturbative analysis of the whole system is presented. In II.F the zeroth order solutions are derived. In II.G the sampled photocurrent at the receiver side is analyzed, and in section II.H the variance of nonlinear distortions are computed. In section III conclusions are drawn. The mathematical background is covered in Appendix A, whereas the detailed derivation of the variance of the SOA-induced nonlinear distortions is presented in Appendix B.    
     
\section{Theory}

\subsection{Basic definitions}

\begin{figure*}[!t]
\centering
\includegraphics[width=7.0in]{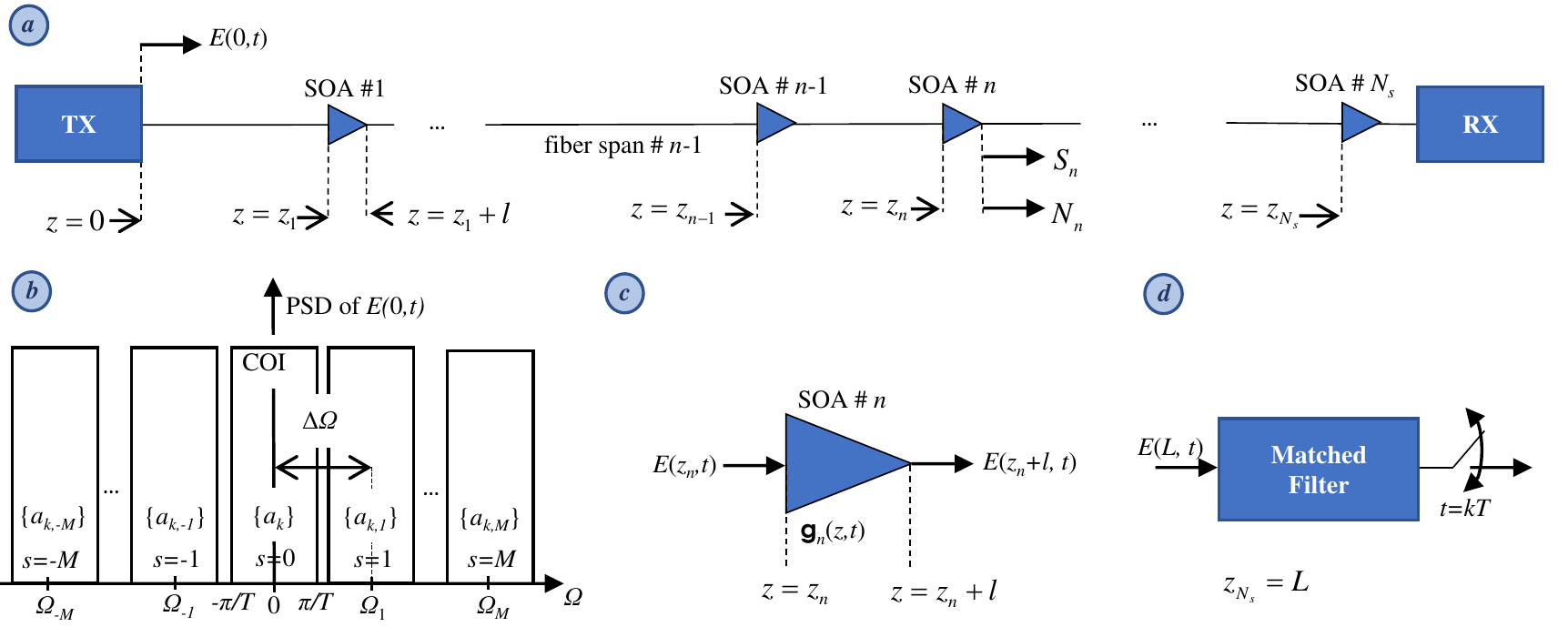}
\hfil
\caption{(a): The fiber-optic multi-span link setup, composed of $N_s$ spans. The $n$'th noisy SOA is placed at the end of the $n$'th span, \emph{i.e.}, at $z=z_n$. The transmitter, TX, is at $z=z_0=0$, and the receiver, RX, is at $z=z_{N_s}$. The total signal (noise) power at the output of the $n$'th SOA is $S_n$ ($N_n$). (b): the power spectral density (PSD) of the total optical field $E(0, t)$ at TX output, COI: channel of interest, $a_{k,s}$ is the $k^{\text{th}}$ information of symbol of the $s^{\text{th}}$ channel, $s=0$ corresponds to COI. $a_{k}$ is the $k^{\text{th}}$ information of symbol of the COI. (c): the  $n^{\text{th}}$ SOA placed at the end of the $n^{\text{th}}$ span, \emph{i.e.}, $z=z_n$. The SOA cavity length is $l$. The material gain of the $n^{\text{th}}$ SOA is $\mathsf g_n(z,t)$ inside the SOA cavity. (d): the ideal coherent RX to receive the COI, composed of the matched filter followed by symbol-spaced sampler.}
\label{fig_soa_multispan_problem_setup} 
\end{figure*}

We consider a multi-span coherent WDM transmission system, with an SOA at the end of each span, as illustrated in Fig.~\ref{fig_soa_multispan_problem_setup}.a. In \cite{Ghazisaeidi2017:JLT} we presented a detailed analysis of the EDFA-based multi-span coherent WDM systems, assuming polarization multiplexed signals, and computed the variance of nonlinear signal-signal (SS) and noise-signal (NS) distortions of the sampled received signals of the channel of interest (COI). In this work, we consider single polarization, replace EDFAs by SOAs, and calculate the variance of SS distortions, using first-order regular perturbation (FRP). We adopt exactly the same notation as in \cite{Ghazisaeidi2017:JLT}, except for variables and parameters related to modeling the optical amplifiers, where the notations for the amplifier gain are more elaborate than \cite{Ghazisaeidi2017:JLT}, and new symbols for the SOA physical parameters are introduced. Table I lists the definition of basic variables and parameters used in this work. If a symbol/notation is not explicitly defined in the text, it should be looked up in Table I. For convenience, we have collected the basic mathematical definitions and identities, required throughout the paper in Appendix \ref{app: preliminaries}.    

The equivalent baseband total optical field envelop at TX output is (cf. Table I and Fig.~\ref{fig_soa_multispan_problem_setup})
\begin{multline} \label{eq: E(0,t) def.}
E\left( {0,t} \right) = \sum\limits_k {{a_k}{A_k}\left( {0,t} \right)}  + \\
\sum\limits_{k,s \ne 0} {{a_{k,s}}{A_{k,s}}\left( {0,t - \delta {T_s}} \right)\exp \left[ { - i{\Omega _s}t + i{\phi _s}\left( 0 \right)} \right]}. 	
\end{multline}
Note that $A_k(z,t)=A_0(z,t-kT)$. We assume the pulse shape of all WDM channels is the ideal Nyquist pulse, with roll-off 0 and energy $\mathcal E=PT$, \emph{i.e.}, 
\begin{equation}
\mathcal{E} = \int_{ - \infty }^{ + \infty } {{{\left| {{A_0}\left( {0,t} \right)} \right|}^2}dt}. 	
\end{equation}
We will sometimes work with the power normalized pulse-shape $U_0(z,t)=A_0(z,t)/\sqrt\mathcal{E}$ in the sequel.
\begin{table*}[!t]
\renewcommand{\arraystretch}{1.0}
\caption{Definition of the basic symbols and notations}
\label{tab: simulation params.}
\centering
\begin{tabular}{|c||c||c||c|}
\hline
$c$      &  speed of light [$m/s$] & $\alpha$ & fiber loss coefficient [$1/m$] \\
\hline
$\hbar$  & Planck constant divided by $2\pi$ [$Js$] & $\beta_2$ & fiber group velocity dispersion  coefficient [$s^2/m$] \\
\hline
$t$      & time [$s$] & $\gamma$ & fiber Kerr nonlinear coefficient [$1/W/m$] \\
\hline
$f$     & frequency [Hz] & $n_{\text{sp}}$ & SOA population inversion factor \\
\hline
$\omega$     & angular frequency [$\text{rad.}/s$] & $P_{sat}$ & SOA saturation power [$W$] \\
\hline 
$\omega_0$   & center angular frequency [$\text{rad.}/s$] & $\tau_c$ & SOA carrier lifetime [$s$] \\
\hline
$\lambda_0$  & center wavelength [$nm$] & $\alpha_H$ & SOA linewidth enhancement factor \\
\hline
$z$     & spatial coordinate along propagation direction [$m$] & $l$ & cavity length of the SOA [$m$] \\
\hline
$z_0$   & coordinate of the TX & $\mathsf g_o$ & small-signal material gain coefficient of the SOA  \\
\hline
 $z_n$  & coordinate of the end of the $n^{\text{th}}$ span [$m$] & $g_o$ & small-signal integrated gain coefficient of the SOA \\
\hline
$\Delta L$  & span length [$m$] & $\mathsf g_n(z,t)$ & small-signal material gain of the $n^{\text{th}}$ SOA [$1/m$] \\
\hline
$L$    & total transmission distance [$m$]  & $g_n(t)$ & integrated gain coefficient of the $n^{\text{th}}$ SOA \\
\hline
$N_s$  & number of spans & $\mathcal I_k$ & $k^{\text{th}}$ sample of matched filtered photocurrent \\
\hline
COI  & channel of interest & $b_k$ & nonlinear signal-signal distortion of the $k^{\text{th}}$ received symbol \\
\hline
$s$    & WDM channels discrete index & $n_k$ &  ASE noise samples of the $k^{\text{th}}$ received symbol of COI \\
\hline
$M$    & number of WDM channels on each side of the COI  & $u_f(t)$ & impulse response of the matched filter \\
\hline
$\Delta\Omega$  & channel spacing [${\text{rad.}}/s$] & $E(z,t)$ & optical field [$\sqrt W$] \\
\hline
$\Omega_s$      & base-band center frequency of the $s^{\text{th}}$ WDM channel & $A_0(z,t)$ & pulse-shape at TX  [$\sqrt W$]\\
\hline
$\delta T_s$    & time shift between COI and the $s^{\text{th}}$ channel at TX & $U(z,t)$ & power normalized optical field  [$\sqrt W$]\\
\hline
$\phi_s(0)$     & time shift between COI and the $s^{\text{th}}$ channel at TX & $U_0(z,t)$ & power normalized pulse-shape at TX  [$\sqrt W$]\\
\hline
$\mathcal E$    & pulse energy [$J$] & $n(z,t)$ & ASE noise process  [$\sqrt W$]\\
\hline
$P$             & channel average power [$W$] & ${}\dag$ & Hermitian conjugation \\
\hline
$T$             & symbol duration [$s$] & ${}^*$ & complex conjugation \\
\hline
$k,m,n,p$       & various discrete indices & $\langle\rangle$ & time and/or ensemble average \\
\hline
$a_k$           & $k^{\text{th}}$ information symbol of the COI & $\langle,\rangle$ & time and/or frequency cross-correlation operator. cf. App.\ref{app: preliminaries} \\
\hline
$a_{k,s}$       & $k^{\text{th}}$ information symbol of the  $s^{\text{th}}$ WDM channel & $\delta(\cdot)$ & Dirac delta function \\
\hline
$\mu_n$         & $n^{\text{th}}$ moment of the constellation & $\theta(\cdot)$ & Heaviside unit step function \\
\hline
$S_n$           & signal power launched into the $n^{\text{th}}$ span & $\delta_{mn}$ & Kronecker delta function \\
\hline
$N_n$           & output noise power generated by the $n^{\text{th}}$ SOA & ${\hat {\mathcal{D}}_{z}}$ & dispersion operator corresponding to a $z$ meter fiber. cf. App. \ref{app: preliminaries} \\
\hline
\end{tabular}
\end{table*}

\subsection{SOA model}

We assume all SOAs are identical components, and adopt the model proposed in \cite{Agrawal1989:JQE}. In this section we suppose that the SOAs are noiseless. The ASE source, will be added later to the propagation equation (cf.(\ref{eq: unique NLSE with gain})). In this work we do not study the nonlinear interaction between signal and ASE noise neither in fiber nor in SOA. The  $n^{\text{th}}$ SOA is placed at the end of the $n^{\text{th}}$ span, and extends from $z=z_n$ to $z=z_n+l$. Inside the SOA waveguide cavity, \emph{i.e.}, for $z_n\leq z\leq z_n+l$, the material gain coefficient of the $n^{\text{th}}$ amplifier, $\mathsf g_n(z,t)$, satisfies the following equation 
\begin{equation}\label{eq: gain rate equation}
\tau_c{\partial_t}\mathsf g_n(z,t) = \mathsf g_o - \mathsf g_n(z,t) - \mathsf g_n(z,t)\frac{{{{\left| {E(z,t)} \right|}^2}}}{{P_{sat}}}.
\end{equation}
In this work, we denote the reciprocal of the saturation power of the SOA by $\epsilon$. We have
\begin{equation}\label{eq: epsilon def.}
\epsilon=\frac{1}{P_{sat}}
\end{equation} 
The field entering the $n^{\text{th}}$ amplifier is $E(z_n,t)$. Inside the $n^{\text{th}}$ amplifier, $z_n\leq z\leq z_n+l$, the following propagation equation holds
\begin{equation}\label{eq: prop. in nth SOA}
\partial_z E(z,t)=\frac{1}{2}(1-i\alpha_H)\mathsf g_n(z,t)E(z,t).
\end{equation}
We define the integrated gain coefficient of the $n^{\text{th}}$ amplifier, $g_n(t)$ as follows
\begin{equation}\label{eq: integrated gain coeff. def.}
g_n(t)=\int_{z_n}^{z_n+l}dz\mathsf g_n(z,t).
\end{equation}
Similarly, the integrated small-signal gain of the $n^{\text{th}}$ SOA is defined as $g_o=\mathsf g_ol$. It is a well-known fact that the $z$~dependence in (\ref{eq: gain rate equation}) and (\ref{eq: prop. in nth SOA}) can be integrated out, to find the following input-output equations for modeling the $n^{\text{th}}$ SOA
\begin{equation}\label{eq: gain dynamics equation of nth SOA}
\tau_c\frac{d}{dt}g_n(t)=g_o-g_n(t)-\epsilon\left[e^{g_n(t)}-1\right]|E(z_n,t)|^2,
\end{equation}
and
\begin{equation}\label{eq: SOA n, in-out fields eq.}
E(z_n+l,t)=\exp\left\{\frac{1}{2}(1-i\alpha_H)g_n(t)\right\}E(z_n,t).
\end{equation}
\subsection{System model}
Signal propagation in each fiber span is modeled by the nonlinear Schr\"odinger equation (NLSE)\cite{Agrawalbook} 
\begin{equation}\label{eq: NLSE}
\partial_zE(z,t)=-\frac{\alpha}{2}E(z,t)-i\frac{\beta_2}{2}\partial_t^2E(z,t)+i\gamma|E(z,t)|^2E(z,t).
\end{equation} 
Note that (\ref{eq: NLSE}) is valid for
\begin{equation}\label{eq: span z condition}
z_{n-1}+l\leq z\leq z_n\text{, for }n=1,...,N_s.
\end{equation}
On the other hand, for $z_{n}\leq z\leq z_n+l$, the propagation of the optical field inside SOA waveguides is modeled by (\ref{eq: prop. in nth SOA}). In fact, the multi-span link is a cascade of $N_s$ nonlinear fiber spans described by the NLSE interleaved with $N_s$ nonlinear SOAs modeled by (\ref{eq: gain dynamics equation of nth SOA}) and (\ref{eq: SOA n, in-out fields eq.}).

Next, we derive a single partial differential equation describing the field propagation in the whole multi-span link. Note that the SOA waveguide cavity length, $l$, is on the order of millimeters, whereas the span length is a few tens of kilometer; therefore, we let $l\to0$, and $\mathsf g_o\to\infty$, but such that the product $\mathsf g_ol$ remains constant and equal to $g_o$. Moreover, we assume that the material gain coefficient of the SOA is a constant function of $z$ inside the SOA waveguide, and is written as
\begin{equation}\label{eq: material g vs. p approx}
\mathsf g_n(z,t)=g_n(t)p_l(z-z_n),
\end{equation}
where,
\begin{equation}\label{eq: p function def.}
p_l(z)=\left\{ {\begin{array}{*{20}{c}}
  1/l&{0 \leq z\leq l} \\ 
  0&{\text{otherwise}} 
\end{array}} \right..
\end{equation} 
Note that when the SOA cavity length tends to zero, we have $\lim\limits_{l\to0}{p_l}(z)=\delta(z)$, where $\delta(\cdot)$ is the Dirac delta function. So, in the $l\to 0$ limit we have
\begin{equation}\label{eq: g_n (z,t) limit}
\lim\limits_{l\to0}\mathsf g_n(z,t)=g_n(t)\delta(z-z_n).
\end{equation}
Now, we define the following overall material gain coefficient function across the whole link from $z=0$ to $z=z_{N_s}$.  
\begin{equation}\label{eq: g(z,t) definition}
\mathsf g(z,t)=\sum\limits_{n=1}^{N_s}g_n(t)\delta(z-z_n). 
\end{equation}
Using (\ref{eq: g(z,t) definition}), we can describe the field propagation from $z=0$ to $z=z_{N_s}$ by a single partial differential equation, which is
\begin{multline}\label{eq: unique NLSE with gain}
\partial_zE(z,t)=\frac{1}{2}\left(1-i\alpha_H\right)\mathsf g(z,t)E(z,t) \\
-\frac{\alpha}{2}E(z,t)-i\frac{\beta_2}{2}\partial_t^2E(z,t)+i\gamma|E(z,t)|^2E(z,t)+n(z,t).
\end{multline} 
In (\ref{eq: unique NLSE with gain}), we have manually added the spatio-temporal ASE process $n(z,t)$. It is a circular complex Gaussian process with zero mean and the following spectral domain autocorrelation
\begin{equation}
	\left\langle{\tilde n}^*(z,\omega)\tilde n(z',\omega')\right\rangle=2\pi\tilde C(z,\omega)\delta(z-z')\delta(\omega-\omega').
\end{equation}
where, we have
\begin{equation} \label{eq: Csw noise in p(z) } 
\tilde C(z,\omega) = \frac{\hbar\omega _0}{2}n_{\text{sp}}\sum\limits_{n = 1}^{{N_s}} {(e^{\bar g_n}-1)\delta(z-z_n)}.
\end{equation}
Note that although the field propagation is described by a single equation, (\ref{eq: unique NLSE with gain}), we still have $N_s$ reservoir equations for the gain dynamics of the SOAs, (cf. (\ref{eq: gain dynamics equation of nth SOA}), and note that $n=1,...,N_s$ ), coupled to the propagation equation (\ref{eq: unique NLSE with gain}). Our next task is to analyze the SOAs' nonlinear gain fluctuations in term of perturbation series in $\epsilon$, and derive a single propagation equation that includes the gain dynamics of all the amplifiers as well. 

\subsection{Perturbation analysis of gain dynamics}
Let's write the integrated gain coefficient of the $n^{\text{th}}$ SOA as
\begin{equation}\label{eq: g_n vs. gbar and delta g_n}
g_n(t)=\bar g_n+\delta g_n(t),
\end{equation}
where, $\bar g_n$ is the time averaged integrated gain coefficient, \emph{i.e.},
\begin{equation}\label{eq: gbar def.}
\bar g_n=\left\langle g_n(t)\right\rangle,
\end{equation}
and $\delta g_n(t)$ is the zero-mean integrated gain fluctuations. In order to obtain the equation governing $\bar g_n$, we compute the time average of both sides of (\ref{eq: gain dynamics equation of nth SOA}). We obtain
\begin{equation}\label{eq: gbar_n equation}
0=g_o-\bar g_n-\epsilon\left\langle\left(e^{\bar g_n}e^{\delta g_n(t)}-1\right)|E(z_n,t)|^2\right\rangle.
\end{equation} 
We keep only first-order fluctuations of gains and optical field intensity in the analysis. Based on this approximation we have
\begin{equation}\label{eq: edg_n = 1 + dg_n}
e^{\delta g_n(t)}\cong 1+\delta g_n(t).
\end{equation}
We also assume the second-order fluctuations $\langle\delta g_n(t)\delta|E(z_n,t)|^2\rangle$ are negligible, thus
\begin{equation}\label{eq: g_n E2 uncorrelated}
\langle\delta g_n(t)|E(z_n,t)|^2\rangle=\langle \delta g_n(t)\rangle\langle |E(z_n,t)|^2\rangle=0.
\end{equation}
Using (\ref{eq: edg_n = 1 + dg_n}) and (\ref{eq: g_n E2 uncorrelated}), we find the following equation for the average gain of the $n^{\text{th}}$ SOA
\begin{equation}\label{eq: static gain equation}
0=g_o-\bar g_n-\epsilon\left(e^{\bar g_n}-1\right)\left\langle|E(z_n,t)|^2\right\rangle.
\end{equation}
We neglect signal depletion by ASE noise, and assume that average gain perfectly compensates the span loss. In other words,  
\begin{equation}\label{eq: gbar n definition}
\bar g_n=\alpha\Delta L.
\end{equation} 
Let's denote the power span loss by $\eta$, which is
\begin{equation}\label{eq: eta def.}
\eta  = e^{-\alpha\Delta L}.
\end{equation}
Note that
\begin{equation}\label{eq: E average}
\left\langle|E(z_n,t)|^2\right\rangle=\eta P_{tot}.
\end{equation}
The $P_{tot}$ is the total power at the output of each SOA. We have $P_{tot}=N_{ch}P$, where $N_{ch}$ is the number of WDM channels. In this work, we assume $N_{ch}$, is an odd number, $N_{ch}= 2M+1$, and that the COI is the $(M+1)^{\text{th}}$ channel located at the center of the WDM grid (cf. Fig. \ref{fig_soa_multispan_problem_setup}.b). 
Using (\ref{eq: edg_n = 1 + dg_n})-(\ref{eq: E average}), the (\ref{eq: gbar_n equation}) results in
\begin{equation}\label{eq: static gain power equation}
\alpha\Delta L+\epsilon(1-\eta)P_{tot}=g_o.
\end{equation} 
Now, using (\ref{eq: gain dynamics equation of nth SOA}), (\ref{eq: g_n vs. gbar and delta g_n}), (\ref{eq: static gain equation}), and (\ref{eq: static gain power equation}), we find the following equation for the gain fluctuations
\begin{multline}\label{eq: equation for dg_n}
\tau_c\frac{d}{dt}\delta g_n(t)+\delta g_n(t) = \epsilon(1-\eta)P_{tot}\\
-\epsilon e^{\bar g_n}\delta g_n(t)|E(z_n,t)|^2-\epsilon(e^{\bar g_n}-1)|E(z_n,t)|^2.
\end{multline}
Now we expand the integrated gain coefficient zero-mean fluctuations and the electrical field in terms of regular perturbation series with respect to the parameter $\epsilon$, as  
\begin{equation}\label{eq: dg_n perturbative series in epsilon}
\delta g_n(t)=\sum\limits_{m=0}^\infty\epsilon^m\delta g_n^{(m)}(t),
\end{equation} 
and
\begin{equation}\label{eq:E(z,t) perturbative fluctuations}
E(z,t)=\sum\limits_{m=0}^\infty\epsilon^mE^{(m)}(z,t).
\end{equation}
In (\ref{eq: dg_n perturbative series in epsilon}) and (\ref{eq:E(z,t) perturbative fluctuations}) $\delta g_n^{(m)}$, and $E^{(m)}$, are the $m^{\text{th}}$ order perturbation corrections to the integrated gain coefficient and the optical field respectively. In the sequel we will only compute terms up to the first order in $\epsilon$. The FRP approximations to the integrated gain and the optical field are 
\begin{equation}\label{eq: g_n,FRP def.}
\delta g_{n,FRP}(t)=\delta g_n^{(0)}(t)+\epsilon\delta g_n^{(1)}(t),
\end{equation}
and
\begin{equation}\label{eq: E_FRP def.}
E_{FRP}(z,t)=E^{(0)}(z,t)+\epsilon E^{(1)}(z,t).
\end{equation} 
Note that
\begin{equation}\label{eq: dg_n vs.fg_n,FRP}
\delta g_n(t)=\delta g_{n,FRP}(t)+ \mathcal O(\epsilon^2),
\end{equation}
and
\begin{equation}\label{eq: E vs. E_FRP}
E(z,t)=E_{FRP}(z,t)+\mathcal O(\epsilon^2).
\end{equation}
Now we substitute (\ref{eq: dg_n perturbative series in epsilon}), and (\ref{eq:E(z,t) perturbative fluctuations}) into (\ref{eq: equation for dg_n}), and extract the equations for zeroth and first order perturbations. The equation governing the zeroth-order integrated gain fluctuations is
\begin{equation}\label{eq:dg_n zeroth order}
\tau_c\frac{d}{dt}\delta g_n^{(0)}(t)+\delta g_n^{(0)}(t)=0.
\end{equation} 
Similarly, we derive the following equation for the first-order perturbation correction to the integrated gain coefficient
\begin{multline}\label{eq: dg_n first_order}
\tau_c\frac{d}{dt}\delta g_n^{(1)}(t)+\delta g_n^{(1)}(t)=(1-\eta)P_{tot}\\
-e^{\bar g_n}\delta g_n^{(0)}(t)|E^{(0)}(z_n,t)|^2-(e^{\bar g_n}-1)|E^{(0)}(z_n,t)|^2.
\end{multline}
We suppose that the system is turned on in $t=-\infty$, and that transient solutions of (\ref{eq:dg_n zeroth order}) and (\ref{eq: dg_n first_order}) are tended to zero. Based on this assumption we have
\begin{equation}\label{eq: dg_n^0 solution}
\delta g_n^{(0)}(t) = 0.
\end{equation}
Now we substitute (\ref{eq: dg_n^0 solution}) into (\ref{eq: dg_n first_order}), and solve (\ref{eq: dg_n first_order}) to obtain
\begin{multline}\label{eq: dg_n^1}
\delta g_n^{(1)}(t)=(1-\eta)P_{tot}\\
-(e^{\bar g_n}-1)\int_{-\infty}^{+\infty} d\tau r_{c}(\tau)|E^{(0)}(z_n,t-\tau)|^2.
\end{multline} 
The $r_{c}(\cdot)$ is the carrier density pulsation (CDP) impulse response of the $n^{\text{th}}$ SOA carrier reservoir. It is explicitly written as
\begin{equation}\label{eq: r_cdp def.}
{r_{c}}\left( t \right) = \frac{1}{{{\tau _c}}}{e^{{{ - t} \mathord{\left/
 {\vphantom {{ - t} {{\tau _c}}}} \right.
 \kern-\nulldelimiterspace} {{\tau _c}}}}}\theta \left( t \right).
\end{equation}
The function $\theta(\cdot)$ is the Heaviside unit step function. Note that $\int_{-\infty}^{+\infty}r_c(t)dt=1$; therefore, as a sanity check, we calculate $\langle\delta g_n^{(1)}(t)\rangle=0$.
 
We substitute (\ref{eq: g_n vs. gbar and delta g_n}), into (\ref{eq: g(z,t) definition}), and derive the following expression for the FRP approximation of the overall material gain coefficient 
\begin{equation}\label{eq: overall material gain coeff. FRP}
\mathsf g_{FRP}(z,t)=\mathsf g^{(0)}(z)+\epsilon\mathsf g^{(1)}(z,t),
\end{equation}
where,
\begin{equation}\label{eq: overall g_o(z) def.}
\mathsf g^{(0)}(z)=\sum\limits_{n=1}^{N_s}\bar g_n\delta(z-z_n). 
\end{equation}
\begin{equation}\label{eq: overall g_n^1(z) def}
\mathsf g^{(1)}(z,t)=-\mathsf d(z)\int_{-\infty}^{+\infty}d\tau r_{c}(\tau)|E^{(0)}(z,t-\tau)|^2,
\end{equation} 
where
\begin{equation}\label{eq: d(z) def.}
\mathsf d(z) = \left(e^{\bar g_n}-1\right)\sum\limits_{n=1}^{N_s}\delta(z-z_n).
\end{equation}
In writing (\ref{eq: overall g_n^1(z) def}) we used the \emph{sifting} property of Dirac delta function, \emph{i.e.},
\begin{equation}\label{eq: Dirac sifting property}
|E^{(0)}(z_n,t)|^2\delta(z-z_n)=|E^{(0)}(z,t)|^2\delta(z-z_n).
\end{equation} 
From now on, we work with $\mathsf g_{FRP}$ instead of $\mathsf g$. We substitute (\ref{eq: overall material gain coeff. FRP}) into (\ref{eq: unique NLSE with gain}), and obtain an FRP approximation to the NLSE (FNLSE), which is
\begin{multline}\label{eq: FNLSE}
\partial_z E(z,t)=\frac{1}{2}(1-i\alpha_H)\mathsf g^{(0)}(z)E(z,t)\\
+\epsilon\frac{1}{2}(1-i\alpha_H)\mathsf g^{(1)}(z,t)E(z,t)\\
-\frac{\alpha}{2}E(z,t)-i\frac{\beta_2}{2}\partial_t^2E(z,t)+i\gamma|E(z,t)|^2E(z,t)\\
\end{multline} 
 At this point, we define the power normalized optical field $U(z,t)$ by the following equation
\begin{multline}\label{eq: U(z,t) def.}
E(z,t)=\\
\sqrt\mathcal E\exp\left\{\frac{1}{2}(1-i\alpha_H)\int_0^z dz'\mathsf g^{(0)}(z')-\frac{\alpha z}{2}\right\}U(z,t).
\end{multline} 
The Perturbation series of the normalized field is
\begin{equation}\label{eq:U perturbation series}
U(z,t)=\sum\limits_{m=0}^\infty{\epsilon^m}U^{(m)}(z,t).
\end{equation} 
The noarmalization relation, (\ref{eq: U(z,t) def.}), is extended to all orders of perturbation $m=0,...,\infty$, \emph{i.e.},
\begin{multline}\label{eq: U^m(z,t) def.}
E^{(m)}(z,t)=\\
\sqrt\mathcal E\exp\left\{\frac{1}{2}(1-i\alpha_H)\int_0^z dz'\mathsf g^{(0)}(z')-\frac{\alpha z}{2}\right\}U^{(m)}(z,t).
\end{multline}
After substituting (\ref{eq: U(z,t) def.}) into (\ref{eq: FNLSE}) and simplification, we obtain the following normalized FNLSE
\begin{multline}\label{eq: Normalized FNLSE}
\partial_z U(z,t)=-i\frac{\beta_2}{2}\partial_t^2U(z,t)+i\gamma\mathcal Ef(z)|U(z,t)|^2U(z,t)\\
-\epsilon \mathcal E\frac{1}{2}\left(1-i\alpha_H\right)\mathsf d(z)f(z)\times\\
\int_{-\infty}^{+\infty} d\tau r_c(\tau)|U^{(0)}(z,t-\tau)|^2 U(z,t)+\frac{n(z,t)}{\sqrt{\mathcal Ef(z)}}.
\end{multline}
The function $f(z)$ is 
\begin{equation}\label{eq: f(z) def.}
\begin{array}{*{20}{c}}
 f(z)=e^{-\alpha(z-z_{n-1})},&z_{n-1}\leq z\leq z_n,n = 1,...,{N_s}.
\end{array}
\end{equation} 
Equation (\ref{eq: Normalized FNLSE}) is the main result of this section. We can describe the whole system by a single equation at the cost of treating the SOA gain fluctuations as first order perturbations in terms of the reciprocal of SOA saturation power, denoted by $\epsilon$. The first and second terms on the right hand side of (\ref{eq: Normalized FNLSE}) correspond to the fiber group velocity dispersion (GVD) and fiber Kerr nonlinearity. If the amplifiers are linear, $\epsilon=0$, and $g_n(t)=\alpha\Delta L$; otherwise, the third term models the nonlinear distortions induced by the cascade of the SOAs as the optical field propagates along the multi-span link. In the next section, we present the FRP analysis of the normalized FNLSE, (\ref{eq: Normalized FNLSE}), in terms of both $\epsilon$ and $\gamma$.
\subsection{Perturbation analysis of the normalized FNLSE}
 We substitute the perturbation series of the power normalized optical field, $U(z,t)$, in terms of the SOA nonlinear parameter $\epsilon$, \emph{i.e.}, equation (\ref{eq:U perturbation series}) into (\ref{eq: Normalized FNLSE}). The equation for the zeroth order term, $U^{(0)}(z,t)$, is
\begin{multline}\label{eq: U^0 equation}
\partial_z U^{(0)}(z,t)=\\
-i\frac{\beta_2}{2}\partial_t^2 U^{(0)}(z,t)+i\gamma\mathcal Ef(z)|U^{(0)}(z,t)|^2U^{(0)}(z,t).
\end{multline} 
We consider the perturbation series of $U^{(0)}(z,t)$ in terms of the fiber nonlinear coefficient $\gamma$
\begin{equation}\label{eq: U^0 in terms of gamma}
U^{(0)}(z,t)=u^{(0,0)}(z,t)+\gamma u^{(0,1)}(z,t)+\mathcal O(\gamma^2).
\end{equation}
We substitute (\ref{eq: U^0 in terms of gamma}) into (\ref{eq: U^0 equation}) and obtain the following equations for the zeroth-order and first-order perturbation series (with respect to $\gamma$) terms of $U^{(0)}$. For the zeroth-order term we have 
\begin{equation}\label{eq: u^00 equation}
\partial_zu^{(0,0)}(z,t)=-i\frac{\beta_2}{2}\partial_t^2u^{(0,0)}(z,t)+\frac{n(z,t)}{\sqrt{\mathcal Ef(z)}},
\end{equation} 
and, for the first-order term we have
\begin{multline}\label{eq:u^01 equation}
\partial_zu^{(0,1)}(z,t)=\\
-i\frac{\beta_2}{2}\partial_t^2u^{(0,1)}(z,t)+i\gamma\mathcal Ef(z)|u^{(0,0)}(z,t)|^2u^{(0,0)}(z,t).
\end{multline} 
In this work, we ignore all the second-order perturbation terms, \emph{i.e.}, the terms of the order of $\mathcal O(\epsilon^2)$, $\mathcal O(\gamma^2)$, \emph{and} $\mathcal O(\epsilon\gamma)$. Adopting this approximation, the equation for the first-order perturbation terms with respect to $\epsilon$ is
\begin{multline}\label{eq: U^{1} equation}
\partial_zU^{(1)}(z,t)=\\
-i\frac{\beta_2}{2}\partial_t^2U^{(1)}(z,t)-\epsilon\mathcal E\frac{1}{2}(1-i\alpha_H)f(z)\mathsf d(z)\times\\
\int_{-\infty}^{+\infty} d\tau r_c( \tau)|u^{(0,0)}(z,t-\tau)|^2u^{(0,0)}(z,t).
\end{multline} 

\subsection{Solutions of the zeroth-order and first-order equations}
The solution of (\ref{eq: u^00 equation}) is
\begin{multline} \label{eq: zeorth-order complete solution detailed}
u^{(0,0)}(z,t)=\sum\limits_k a_ku_k^{(0,0)}(z,t)  + \\
 \sum\limits_{k,s \ne 0} a_{k,s}u_{k,s}^{(0,0)}(z,t-t_s(z))
\exp\left[-i\Omega_st+i\phi_s(z)\right]+\\
{{\hat {\mathcal{D}}}_z}\left[ {{u_{{\rm{ASE}}}}\left( {z,t} \right)} \right],
\end{multline}
where\footnote{cf. \cite{Mecozzi2012:JLT2012} and the discussion leading to Eq.~(72) in \cite{Ghazisaeidi2017:JLT}.}  
\begin{equation} \label{eq: u_ASE def.}
u_{\rm{ASE}}(z,t)=\int_0^z dz'\frac{n(z',t)}{\sqrt{\mathcal{E}f(z')}}.
\end{equation}
The zeroth-order dispersed pulse of the $k^{\text{th}}$ symbol of the $s^{\text{th}}$ channel after propagating up to distance $z$ is
\begin{equation} \label{eq:}
u_{k,s}^{(0,0)}(z,t)=\hat{\mathcal D}_z\left[u_{k,s}^{(0,0)}(0,t)\right].
\end{equation}
The same expression holds for COI, albeit with dropping the subscript $s$. The $u_0^{(0,0)}(0,t)$ is the unperturbed pulse-shape of the COI, which is
\begin{equation}\label{eq: Nyquist def.}
u_0^{(0,0)}(0,t)=\frac{1}{\sqrt T}\text{sinc}\left(\frac{t}{T}\right).
\end{equation} 
All the adjacent channels use the same pulse-shape as the COI. The time shift between COI and the $s^{\text{th}}$ channel, after propagation up to distance $z$ is
\begin{equation} \label{eq: ts(z) def.}
t_s(z)=\delta T_s+\beta_2\Omega_sz.
\end{equation}
The relative phase shift between COI and the $s^{\text{th}}$ channel, after propagation up to distance $z$ is
\begin{equation} \label{eq: phi_s(z) def.}
\phi_s(z)=\phi_s(0)+\frac{\beta_2\Omega_s^2z}{2}.
\end{equation}
We suppose $\delta T_s=0$, $\phi_s(0)=0$, and $\Omega_s = s\Delta \Omega$, in the rest of the work.
 
The solution of (\ref{eq:u^01 equation}) is presented in Eq.~(90) of \cite{Ghazisaeidi2017:JLT} (cf. also \cite{Mecozzi2012:JLT2012} and \cite{Dar2015:JLT2015}), and will not be repeated here. The solution of (\ref{eq: U^{1} equation}) is  
\begin{multline}\label{eq: U^1 solution}
U^{(1)}(z,t)=\epsilon\mathcal E\frac{1}{2}(1-i\alpha_H)\hat{\mathcal D}_z\int_0^z dz'\mathsf d(z')f(z')\times\\
\hat{\mathcal D}_{z'}^\dag\left[\int_{-\infty}^{+\infty} d\tau r_c(\tau)|u^{(0,0)}(z',t-\tau)|^2u^{(0,0)}(z',t)\right].
\end{multline}
\subsection{Sampled photocurrent analysis}
The ideal coherent receiver in this work consists of a matched filter to the COI pulse-shape followed by ideal symbol-spaced sampling. cf. Fig.~\ref{fig_soa_multispan_problem_setup}.c. The impulse response of the matched filter, $u_f$, is given by  
\begin{equation}\label{eq: u_f def.}
u_f^*(-t)=\hat {\mathcal D}_L\left[u_0^{(0,0)}(0,t)\right],
\end{equation}
The sampled photocurrent $\mathcal I_k$ is given by the following equation
\begin{equation} \label{eq: I_k xcorr developed}
\mathcal I_k=\left\langle\hat {\mathcal D}_L\left[ u_k^{(0,0)}(0,t)\right], U(L,t)\right\rangle.
\end{equation}
The FRP approximation to $\mathcal I_k$ is denoted by $\mathcal J_k$. We have
\begin{equation}\label{eq:I_k vs. J_k}
\mathcal I_k=\mathcal J_k+\mathcal O(\epsilon^2)+\mathcal O(\gamma^2)+\mathcal O(\epsilon\gamma).
\end{equation} 
From now on, we study only the  $\mathcal J_k$. It is given by
\begin{equation} \label{eq: J_k xcorr developed}
\mathcal J_k=\left\langle\hat {\mathcal D}_L\left[ u_k^{(0,0)}(0,t)\right], U_{FRP}(L,t)\right\rangle,
\end{equation}
where
\begin{equation}\label{eq: UFRP}
U_{FRP}(z,t)=u^{(0,0)}(z,t)+\gamma u^{(0,1)}(z,t)+\epsilon U^{(1)}(z,t).
\end{equation} 
We substitute (\ref{eq: UFRP}) into (\ref{eq: J_k xcorr developed}), and we obtain
\begin{equation}\label{eq:J_k analysis}
\mathcal J_k=a_k+b_k+b'_k+n_k,
\end{equation}     
where $a_k$ is the $k^{\text{th}}$ symbol of the COI, $b_k$ is the total fiber-induced SS nonlinear distortion, and $b'_k$ is the total SOA-induced SS nonlinear distortions. The detailed analysis of $b_k$, using exactly the same notation as this work, is presented in \cite{Ghazisaeidi2017:JLT} (cf. also \cite{Mecozzi2012:JLT2012} and \cite{Dar2015:JLT2015}). Here, we focus on the SOA-induced nonlinear distortions. We have
\begin{equation}\label{eq:b'_k definition}
b'_k=\left\langle\hat {\mathcal D}_L\left[ u_k^{(0,0)}(0,t)\right], U^{(1)}(L,t)\right\rangle.
\end{equation} 
Now we substitute (\ref{eq: U^1 solution}) into (\ref{eq:b'_k definition}) and use the dispersion exchange formula (\ref{eq: DEF def.}),\cite{Mecozzi2012:JLT2012,Ghazisaeidi2017:JLT}. We obtain
\begin{multline}\label{eq:b'_k detailed bracket}
b'_k=\epsilon\mathcal E\frac{1}{2}(1-i\alpha_H)\int_0^L dz\mathsf d(z)f(z)\int_{-\infty}^{+\infty} dtu_k^{(0,0)}(z,t)\times\\
\int_{-\infty}^{+\infty} d\tau r_c(\tau)|u^{(0,0)}(z,t-\tau)|^2u^{(0,0)}(z,t).
\end{multline} 
\subsection{Computing the variance of nonlinear distortions}
In this section we give the expressions for the variance of the terms on the right hand side of (\ref{eq:J_k analysis}). The variance of the ASE, $\sigma_{ASE}^2=\left\langle|n_k|^2\right\rangle$ is (cf. Eq.~(98) in \cite{Ghazisaeidi2017:JLT})
\begin{equation}\label{eq:}
\sigma _{ASE}^2(P)=\frac{\hbar\omega_0}{2PT}n_{\text{sp}}\left(e^{\bar g}-1\right)N_s.
\end{equation} 
The variance of single polarization signal-signal fiber-induced total nonlinear distortions, $\sigma_{SS,f}^2=\left\langle|b_k|^2\right\rangle$, is given by  (cf. Eq.~(105) in \cite{Ghazisaeidi2017:JLT}) 
\begin{multline} \label{eq: sig^2_SS,f variance}
\sigma_{SS,f}^2(P)=\\
\gamma^2P^2\{2\mathcal{X}_1 + (\frac{\mu_4}{\mu_2^2}-2) [\mathcal{X}_2 + 4\mathcal{X}_3 + 4\mathcal{X}_4] + \\
(\frac{\mu_6}{\mu_2^3}-9\frac{\mu_4}{\mu_2^2} + 12)\mathcal{X}_5 + 4\sum\limits_s[\mathcal{X}_{1,s}+(\frac{\mu_4}{\mu_2^2}-2)\mathcal{X}_{3,s}] + \\
\sum\limits_s \sum\limits_{s'} \mathcal{X}_{1,s,s'}\},
\end{multline}
where the $\mu_n$ is the $n^{\text{th}}$ moment of the constellations, and the various $\mathcal X$ coefficients on the right hand side of (\ref{eq: sig^2_SS,f variance}) are given by Eqs. (202)-(208) in \cite{Ghazisaeidi2017:JLT}.

The new contribution of this work, is calculating the variance of signal-signal SOA-induced total nonlinear distortions $\sigma_{SS,s}^2=\left\langle|b'_k|^2\right\rangle$. It is given by
\begin{multline} \label{eq: sig^2_SS,s variance}
\sigma_{SS,s}^2(P)=\\
\frac{1}{4}\epsilon^2P^2(1+\alpha_H^2)(\bar G-1)^2\{2\mathcal{Y}_1 + (\frac{\mu_4}{\mu_2^2}-2) [\mathcal{Y}_2 + 4\mathcal{Y}_3 + 4\mathcal{Y}_4] + \\
(\frac{\mu_6}{\mu_2^3}-9\frac{\mu_4}{\mu_2^2} + 12)\mathcal{Y}_5 + 4\sum\limits_s[\mathcal{Y}_{1,s}+(\frac{\mu_4}{\mu_2^2}-2)\mathcal{Y}_{3,s}] + \\
\sum\limits_s \sum\limits_{s'} \mathcal{Y}_{1,s,s'}\},
\end{multline}
where $\bar G=e^{\bar g}$, and the $\mathcal Y$ coefficients on the right hand side of (\ref{eq: sig^2_SS,s variance}) are computed in the Appendix.~\ref{app: SOA induced nonlinearity}.
%
\section{Conclusions}
We presented for the first time, a theory of signal propagation in multi-span wavelength division multiplexed coherent transmission systems, which use in-line semiconductor optical amplifiers (SOA). The SOAs were modeled by a standard reservoir model. The reciprocal of the SOA saturation power was treated as a new second perturbation parameter, besides the fiber nonlinear coefficient. First, an equivalent nonlinear Schr\"odinger equation, capturing both signal propagation in fiber and the SOAs gain dynamics was derived. Then time-domain first-order regular perturbation theory was applied to this equation, and the signal-to-noise ratio of the sampled received photocurrent was computed.  

\appendix
\subsection{Notations and preliminary material}\label{app: preliminaries}
The Fourier transform pair in this work is
\begin{equation} \label{eq: FT}
\tilde x\left( {z,\omega } \right) = \mathcal{F} \left[ {x\left( {z,t} \right)} \right] = \int_{ - \infty }^{ + \infty } {dt\exp \left( {i\omega t} \right)x\left( {z,t} \right),} 	
\end{equation}
 \begin{equation} \label{eq: IFT}
	x\left( {z,t} \right) = {\mathcal{F}^{ - 1}}\left[ {\tilde x\left( {z,\omega } \right)} \right] = 
	\frac{1}{{2\pi }}\int_{ - \infty }^{ + \infty } {d\omega\exp \left( { - i\omega t} \right)\tilde x\left( {z,\omega } \right),} 
\end{equation}
where, $\mathcal{F}$ stands for the Fourier transform, $\mathcal{F}^{-1}$ stands for the inverse Fourier transform, $\omega=2\pi f$ is the angular frequency, and $f$ is the frequency. Throughout this paper, a waveform with a tilde on top  is the Fourier transform of the waveform denoted by the same symbol but without tilde. 
We define the following notations for waveforms cross-correlations in time domain
\begin{equation} \label{eq: xcorr td. def.}
	\left\langle {x\left( {z,t} \right),y\left( {z',t'} \right)} \right\rangle  = \int_{ - \infty }^{ + \infty } {d\tau} {x^*}\left( {z,t+\tau} \right)y\left( {z',t'+\tau} \right),
\end{equation}
and in frequency domain
\begin{equation} \label{eq: xcorr fd. def.}
	\left\langle {\tilde x\left( {z,\omega } \right),\tilde y\left( {z',\omega' } \right)} \right\rangle  = \int_{ - \infty }^{ + \infty } {d\nu} {\tilde x^*}\left( {z,\omega +\nu} \right)\tilde y\left( {z',\omega '+\nu} \right),
\end{equation}
where, the superscript $*$ stands for complex conjugation. 
On the other hand, we use the notation $\left\langle {x\left({z,t} \right)} \right\rangle$ to denote the ensemble average over the space of all sample waveforms of the stochastic process $x(z,t)$. The randomness of the processes in this work is due to information symbols, which are assumed to be independent identically distributed (i.i.d.) discrete random variables with phase-isotropic distributions, and also due to amplifier noise. In our notation, the Parseval's theorem is stated as follows
\begin{equation} \label{eq: Parseval theorem}
\left\langle {x\left( {z,t} \right),y\left( {z',t} \right)} \right\rangle  = \frac{1}{{2\pi }}\left\langle {\tilde x\left( {z,\omega } \right),\tilde y\left( {z',\omega } \right)} \right\rangle. 	
\end{equation}

We introduce the following notation for the dispersion operator in the frequency domain
\begin{equation} \label{eq: dispersion op. def. fd}
{\hat {\mathcal{D}}_{z'}}\left[ {\tilde x\left( {z,\omega } \right)} \right] = {\exp \left( {i\frac{{{\omega ^2}}}{2}} {\int_0}^{z'} dz''{\beta _2}(z'') \right)}\tilde x\left( {z,\omega } \right),
\end{equation}
and in the time domain
\begin{multline} \label{eq: dispersion op. def. td}
{\hat {\mathcal{D}}_{z'}}\left[ {x\left( {z,t} \right)} \right] = \\
{\mathcal{F} ^{ - 1}}\left[ {\exp \left( {i\frac{{{\omega ^2}}}{2}} {\int_0}^{z'} dz''{\beta _2}(z'')\right)\tilde x\left( {z,\omega } \right)} \right].
\end{multline}

The following \textit {dispersion exchange formula} (DEF), which is easily proven by Parseval's theorem, will be extensively used in this work (cf. Eqs.(19) and (20) in \cite{Mecozzi2012:JLT2012})
 \begin{equation} \label{eq: DEF def.}
\left\langle {x\left( {z,t} \right),{{\hat {\mathcal{D}}}_{z'}}\left[ {y\left( {z,t} \right)} \right]} \right\rangle  = \left\langle {{{{\hat {\mathcal{D}}}^\dag}_{z'}}\left[ {x\left( {z,t} \right)} \right],y\left( {z,t} \right)} \right\rangle, 	
\end{equation}
where, $\hat{\mathcal D}^{\dag}_{z'}$ is the adjoint of the dispersion operator $\hat{\mathcal{D}}_{z'}$, which is defined to be  
\begin{equation} \label{eq: dispersion op. def. fd adjoint}
{{\hat {\mathcal{D}}^\dag}_{z'}}\left[ {\tilde x\left( {z,\omega } \right)} \right] = {\exp \left( -{i\frac{{{\omega ^2}}}{2}} {\int_{0}}^{z'} dz''{\beta _2}(z'') \right)}\tilde x\left( {z,\omega } \right).
\end{equation}
Finally note that the sinc function is 
\begin{equation}
{\rm{sinc}}\left( x \right) = \frac{{\sin \left( {\pi x} \right)}}{{\pi x}}.
\end{equation}
\subsection{Computing the noise variance of SOA-induced nonlinearity}\label{app: SOA induced nonlinearity}
In this section we outline the computation of the variance of signal-signal SOA-induced nonlinear distortions $\sigma_{SS,s}^2$. We substitute the signal terms on right hand side of (\ref{eq: zeorth-order complete solution detailed}) into (\ref{eq:b'_k detailed bracket}). We ignore nonlinear noise-signals interaction in the SOA in this work. The resulting expression is
\begin{multline} \label{eq: b_k, detailed expression}
{b'_{k}} = \sum\limits_{m,n,p} a_{m+k}a_{n+k}a_{p+k}^*Y_{m,n,p}^{(0,0)}+\\
2\sum\limits_s \sum\limits_{m,n,p}a_{m+k}a_{n+k,s}a_{p+k,s}^*Y_{m,n,p}^{(0,s)}+\\
\sum\limits_{\mathop {s,s'}\limits_{s \ne s'}}\sum\limits_{m,n,p}a_{m+k,s}a_{n+k,s'}a_{p+k,s+s'}^*Y_{m,n,p}^{(s,s')},
\end{multline}
where,
\begin{multline}\label{eq: Y_{m,n,p}^{(s,s')} def}
Y_{m,n,p}^{(s,s')}=\int_{-\infty}^{+\infty}\int_{-\infty}^{+\infty}dtd\tau r_c(\tau)u_0^{(0,0)*}(z,t)\times\\
u_m^{(0,0)}(z,t-t_s-\tau)u_n^{(0,0)}(z,t-t_{s'})\times\\
u_p^{(0,0)*}(z,t-t_{s + s'}-\tau).
\end{multline} 
these coefficients are expressed in the spectral domain as follows
\begin{equation} \label{eq: K^(s,s')_mnp frequency domain}
Y_{m,n,p}^{(s,s')} = \int_{{\rm I\!R}^3} {\frac{{{d^3}\omega }}{{{{\left( {2\pi } \right)}^3}}}} {H'_{{\vec \omega },s,s'}}\left( z \right){e^{i\left( {m{\omega _1} - p{\omega _2} + n{\omega _3}} \right)T}}.
\end{equation}
The symbol $d^n\omega$ stands for $d\omega_1d\omega_2\cdots d\omega_n$ for any positive integer $n$, and the following notational convention is used 
\begin{equation} \label{eq: omega vect. def.}
\vec \omega  = \left( {{\omega _1},{\omega _2},{\omega _3}} \right).
\end{equation}
The integrand function in is 
\begin{multline} \label{eq: H_s,s' def.}
H'_{\vec\omega,s,s'}(z_q)=\Pi_{\vec\omega,s,s'}\mathsf d(z)f(z)\times\\
e^{i\beta_2z[(\omega_2-\omega_3)(\omega_2-\omega_1)-ss'\Delta\Omega^2]}.
\end{multline}
The function $\Pi'_{\vec\omega,s,s'}$ is defined to be
\begin{multline} \label{eq: Pi_omega def.}
\Pi'_{\vec\omega,s,s'} = \tilde r_c(\omega_1-\omega_2-\Omega_{s'})\tilde u_0^{(0)}(0,\omega_1-\Omega_s)\times \\
\tilde u{_0^{(0)*}}(0,\omega_2-\Omega_{s+s'})
\tilde u_0^{(0)}(0,\omega_3-\Omega_{s'})\tilde u{_0^{(0)*}}\left( {0,{\omega_1} - {\omega_2} + {\omega_3}} \right).
\end{multline}
The following notional simplifications are used: $H'_{\vec\omega,s}=H'_{\vec\omega,0,s}$, $\Pi'_{\vec\omega,s}=\Pi'_{\vec\omega,0,s}$, $H'_{\vec\omega}=H'_{\vec\omega,0,0}$, and $\Pi'_{\vec\omega}=\Pi'_{\vec\omega,0,0}$\footnote{Note that the notations $H_{\vec\omega,s,s'}$, and $\Pi_{\vec\omega,s,s'}$ are reserved for the treatment of the $\mathcal X$ coefficients in \cite{Ghazisaeidi2017:JLT}}. 

The rest of derivation follows exactly the steps detailed in the Appendix of \cite{Ghazisaeidi2017:JLT} for the derivation of $\mathcal X$ coefficients. In order to define $\mathcal Y$ coefficients, take Eqs. (134)-(141) of \cite{Ghazisaeidi2017:JLT}, and replace $X$, and $\mathcal X$ by $Y$ and $\mathcal Y$ respectively everywhere in those equations. The next step is writing the integral representations for the $\mathcal Y$ coefficients, which are obtained by taking Eqs. (153)-(160) of \cite{Ghazisaeidi2017:JLT}, and replacing $\mathcal X$ by $\mathcal Y$ and $H$ by $H'$ everywhere in those equations. Finally, we use (\ref{eq: d(z) def.}) and carry out $z$-integrations. The end results are 
\begin{equation} \label{eq: Y_1 final ref.}
\mathcal{Y}_1=\frac{1}{T}\int {\frac{{{d^3}\omega }}{{{{\left( {2\pi } \right)}^3}}}} \left|{\sum\limits_{m = 1}^{{N_s}} {{r'_{\vec \omega ,m}}} } \right|^2,
\end{equation}
\begin{equation} \label{eq: Y_2 final ref.}
\mathcal{Y}_2=\operatorname{Re} \left\{\int {\frac{{{d^4}\omega }}{{{{\left( {2\pi } \right)}^4}}}} \left( {\sum\limits_{m = 1}^{{N_s}} {{r'_{\vec \omega ,m}}} } \right)\left( {\sum\limits_{m = 1}^{{N_s}} {r^{'*}_{\vec \omega ',m}} } \right)\right\},
\end{equation}
\begin{equation} \label{eq: Y_3 final ref.}
\mathcal{Y}_3=\operatorname{Re} \left\{\int {\frac{{{d^4}\omega }}{{{{\left( {2\pi } \right)}^4}}}} \left( {\sum\limits_{m = 1}^{{N_s}} {{r'_{\vec \omega ,m}}} } \right)\left( {\sum\limits_{m = 1}^{{N_s}} {r^{'*}_{\vec \omega'',m}} } \right)\right\},
\end{equation}
\begin{equation} \label{eq: Y_5 final ref.}
\mathcal{Y}_5=T\operatorname{Re} \left\{\int {\frac{{{d^5}\omega }}{{{{\left( {2\pi } \right)}^5}}}} \left( {\sum\limits_{m = 1}^{{N_s}} {{r'_{\vec \omega ,m}}} } \right)\left( {\sum\limits_{m = 1}^{{N_s}} {r^{'*}_{\vec \omega''',m}} } \right)\right\},
\end{equation}
\begin{equation} \label{eq: Y_1_s final ref.}
\mathcal{Y}_{1,s}=\frac{1}{T}\int {\frac{{{d^3}\omega }}{{{{\left( {2\pi } \right)}^3}}}} \left|{\sum\limits_{m = 1}^{{N_s}} {{r'_{\vec \omega,m,s}}} } \right|^2,
\end{equation}
\begin{equation} \label{eq: Y_3_s final ref.}
\mathcal{Y}_{3,s}=\operatorname{Re} \left\{\int {\frac{{{d^4}\omega }}{{{{\left( {2\pi } \right)}^4}}}} \left( {\sum\limits_{m = 1}^{{N_s}} {{r'_{\vec\omega,m,s}}} } \right)\left( {\sum\limits_{m = 1}^{{N_s}} {r^{'*}_{\vec \omega'',m,s}} } \right)\right\},
\end{equation}
\begin{equation} \label{eq: Y_1_s_s' final ref.}
\mathcal{Y}_{1,s,s'}=\frac{1}{T}\int {\frac{{{d^3}\omega }}{{{{\left( {2\pi } \right)}^3}}}} \left|{\sum\limits_{m = 1}^{{N_s}} {{r'_{\vec\omega ,m,s,s'}}} } \right|^2,
\end{equation}
The integrand function in is 
\begin{multline} \label{eq: r_s,s' def.}
r'_{\vec\omega,m,s,s'}=\Pi'_{\vec\omega,s,s'}e^{i\beta_2z_m[(\omega_2-\omega_3)(\omega_2-\omega_1)-ss'\Delta\Omega^2]}.
\end{multline}
Like for $H'$, and $\Pi'$, we adopt the following notational conventions: $r'_{\vec\omega,m,s}=r'_{\vec\omega,m,0,s}$, $r'_{\vec\omega,m}=r'_{\vec\omega,m,0,0}$. The following shorthand notation are used in (\ref{eq: Y_1 final ref.})-(\ref{eq: Y_1_s_s' final ref.}) 
\begin{equation} \label{eq: omega' vect. def.}
\vec \omega ' = \left( {{\omega _4},{\omega _2},{\omega _1} + {\omega _3} - {\omega _4}} \right),
\end{equation}
\begin{equation} \label{eq: omega'' vect. def.}
\vec \omega'' = \left( {{\omega _1},{\omega _4},{\omega _3} - {\omega _2}+{\omega _4}} \right),
\end{equation}
and
\begin{equation} \label{eq: omega''' vect. def.}
\vec \omega'''=(\omega_4,\omega_5,\omega_1-\omega_2+\omega_3-\omega_4+\omega_5).
\end{equation}
The $\omega_i$'s, for $i=1,\cdots,5$, are independent real dummy integration variables.  
%
 


\end{document}